\documentclass[aps,prl,twocolumn,showpacs,floatfix,bibnotes]{revtex4-2}

\usepackage{amssymb,amsmath,amstext}                
\usepackage{graphicx}                                               
\usepackage{epstopdf}                                               
\usepackage{color}                                                     
\usepackage{bm}                                                        
\usepackage{appendix}                                              
\usepackage[utf8]{inputenc}
\usepackage{bbold}
\usepackage{bbm}
\usepackage[normalem]{ulem}
\usepackage{latexsym}
\usepackage{lineno}
\usepackage{xcolor} 
\usepackage{braket}
\usepackage{mathtools}
\usepackage{layouts}
\definecolor{lblue} {RGB}{51,71,158}
\usepackage[colorlinks=true,citecolor=blue,linkcolor=blue,urlcolor=lblue]{hyperref}
\usepackage{lineno}

\newcommand{\kb}{k_\textrm{B}}

\newcommand{\edd}{\epsilon_\textrm{dd}}
\newcommand{\krot}{k_\textrm{r}}
\newcommand{\add}{a_\textrm{dd}}
\newcommand{\etadd}{\eta_\textrm{dd}}
\usepackage{soul}
\usepackage{braket}
\usepackage{upgreek}

\begin{document}

\preprint{APS/123-QED}

\title{Observation of rotonic density enhancement at the boundary of a \\uniform dipolar superfluid
}

\author{Jasmine Kalia}
\thanks{These authors contributed equally.}
\author{Kiryang Kwon}
\thanks{These authors contributed equally.}
\author{William J. Solorio Hernandez}
\thanks{These authors contributed equally.}
\author{Rubaiya R. Emran}
\author{Aryan Prasad}
\author{Richard J. Fletcher}
\email{rfletch@mit.edu}

\affiliation{%
 MIT-Harvard Center for Ultracold Atoms, Research Laboratory of Electronics, and Department of Physics, Massachusetts Institute of Technology, Cambridge, Massachusetts 02139, USA 
}%

\date{\today}
\maketitle

\textbf{
The collective excitations of a quantum liquid lie at the heart of its emergent behaviour~\cite{Nozires:2018}.
In many systems, the variation of excitation energy $\bm{\varepsilon}$ with wavevector $\bm{k}$ exhibits a softening at a characteristic lengthscale, which plays a key role in thermodynamics and transport~\cite{Landau:1941, Landau:1947,Feynman:1954}.
These excitations, termed rotons, often reflect an underlying tendency of the liquid to crystallize into a competing symmetry-broken state~\cite{Nozires:2004}, such as the solid state of helium or the Wigner crystal of electrons in a magnetic field~\cite{Mendez:1983,Lam:1984,Jiang:1990}.
More than $\bm{50}$ years ago, it was predicted that rotons should manifest as emergent density oscillations at the fluid's boundary~\cite{Regge:1972, Rasetti:1978, Dalfovo:1994}, but this effect has not been observed.
Here, we realize an ultracold dipolar superfluid confined in a homogeneous trapping potential and directly confirm the emergence of rotonic density enhancement at its edge. 
This is accompanied by rotonic density fluctuations across the bulk, which are linked to the static edge structure via the fluctuation-dissipation theorem.
Increasing the dipolar character leads to an azimuthal density modulation of the fluid's boundary, which we show to be analogous to the behaviour of a quasi-one-dimensional fluid confined in an annular optical potential.
}

Density oscillations at the boundary of a rotonic liquid occur because the vanishing density imposed by a wall seeds the local organization of neighbouring atoms into periodic planes. 
More quantitatively, softening of $\varepsilon(k)$ at a wavevector $\krot$ implies a corresponding peak in the static density response function $\chi(k)$~\cite{Pitaevskii:2016, Bisset:2019}, which determines the change in the fluid's density arising from a perturbation with wavevector $k$. 
In a seminal work concerning liquid helium~\cite{Regge:1972}, Regge showed that this peak directly manifests as density oscillations at the boundary with spatial period $2\pi/\krot$.
Similar density structures are thought to also accompany quantum vortices~\cite{Chester:1968, Dalfovo:1992, Wilson:2008} and impurities~\cite{Dalfovo:1994b}, and play a key role in dissipation and turbulence~\cite{Amelio:2018}. 
This static density response is closely linked to the ephemeral density fluctuations of the bulk, characterized by the structure factor $S(k)$, as a consequence of the fluctuation-dissipation theorem~\cite{Pitaevskii:2016}.
In fractional quantum Hall liquids, analogous edge effects were first discovered numerically~\cite{Morf:1986,Ciftja:2003} and subsequently associated with the magneto-rotons of the bulk~\cite{Girvin:1986,Kukushkin:2009,Jolad:2010,Wiegmann:2012}. 
More broadly, rotons emerge in numerous disparate systems including neutron stars~\cite{Chamel:2008}, heavy nuclei~\cite{Rasetti:1978}, Rydberg gases~\cite{Henkel:2010}, and moir\'e materials~\cite{Paul:2026}.
However, these density modulations have never been observed.

\begin{figure*}[t]
    \includegraphics[width = \linewidth]{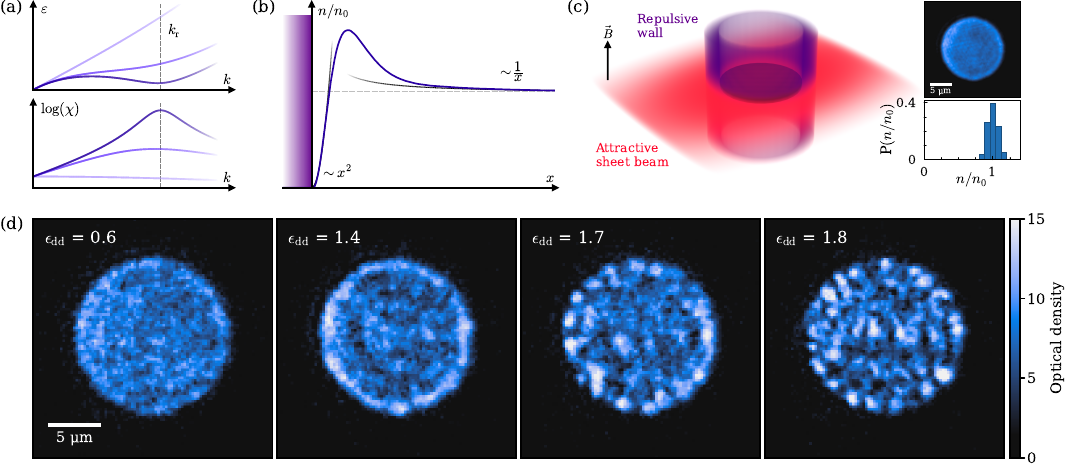}
    \caption{\textbf{Emergent edge density modulation and bulk fluctuations in a uniform dipolar superfluid.} 
    \textbf{a)}~In many quantum fluids, the spectrum of collective excitations $\varepsilon(k)$ features a softening at a characteristic wavevector $\krot$. 
    This softening manifests as a pronounced peak in the static density response function $\chi(k)$, which determines the density modulation arising from an external perturbation.
    The three lines show illustrative excitation spectra and their corresponding response functions.
    \textbf{b)}~In the vicinity of a confining wall (purple), the roton manifests as a modulation of the density $n$ relative to its bulk value $n_0$, reflecting an underlying tendency of the fluid to crystallize. The purple curve is obtained by solving the GP equation and the black lines show analytic functions capturing the short- and long-distance behaviours without any free parameters. The dashed line indicates $n=n_0$. \textbf{c)}~We confine a Bose-Einstein condensate of dipolar $^{166}$Er atoms in a homogeneous disc-shaped trapping potential, formed by an attractive sheet beam (red) intersected by a sharp repulsive wall (purple). An axial magnetic field $\vec{B}$ both aligns the magnetic dipoles and allows control over the $s-$wave scattering length. The inset shows the average profile at a dipolar parameter $\edd=0.6$, and a density histogram indicating a standard deviation of $7\%$. \textbf{d)}~We observe a dramatic evolution of the density profile with increasing rotonic character. Initially, the fluid exhibits a clear density enhancement at the wall, and further increasing $\edd$ results in pronounced azimuthal density modulation along this quasi-one-dimensional boundary. This evolution of the fluid's edge is accompanied by increasing fluctuations at the rotonic lengthscale across the entire bulk. 
    }
    \label{Fig1}
\end{figure*}

Ultracold atomic gases offer an exquisitely controlled platform for exploring quantum fluids. Rotonic phenomena have been engineered by an external drive in spin-orbit coupled gases~\cite{Khamehchi:2014, Ji:2015}, optical cavities~\cite{Mottl:2012,Leonard:2017}, and optical lattices~\cite{Ha:2015}, and in rotating fluids~\cite{Mukherjee:2022}. 
In magnetic dipolar gases, the interplay of long-ranged interatomic forces with the fluid's confinement results in a highly tunable rotonic excitation spectrum~\cite{Santos:2003,Giovanazzi:2004,Ronen:2007}. Rotons have been detected via Bragg spectroscopy~\cite{Petter:2019} and thermal density fluctuations~\cite{Hertkorn:2021}, while tuning their energy to vanish yields a dynamical instability toward spontaneous density modulation~\cite{Chomaz:2018} and the formation of symmetry-broken states, such as droplet crystals and supersolids~\cite{Chomaz:2023,Recati:2023,He:2025, Chandrashekara:2026}. 
However, all prior experiments were performed using inhomogenous gases held in harmonic traps~\cite{Chomaz:2023}, precluding the exploration of boundary physics despite numerous theoretical efforts~\cite{Wilson:2008,Lu:2010,Juhasz:2022,Roccuzzo:2022, Juhasz:2022}.

Here, we realize a Bose-Einstein condensate (BEC) of magnetic $^{166}$Er atoms held in a homogeneous cylindrical optical box potential~\cite{Navon:2021}. 
We observe a pronounced density enhancement at the boundary, whose magnitude grows upon increasing the fluid's rotonic character by tuning either the interparticle interactions or the axial confinement. 
Accompanying this static modulation are rotonic density fluctuations across the fluid's bulk, whose magnitude grows commensurate with the edge density response.
Softening of the roton energy eventually yields a pronounced azimuthal density modulation of the high-density boundary region, which closely mirrors the observed evolution of a quasi-one-dimensional fluid confined in an annular optical trap. 

Before turning to the experiment, we consider the expected edge density profile of a planar dipolar condensate. 
Its rotonic character is encoded in the excitation spectrum~\cite{Santos:2003,Giovanazzi:2004,Baillie:2015}.
For simplicity, we consider the case that only the axial ground state is occupied~\cite{methods},
\begin{equation}
    \varepsilon(k)=\hbar\omega_z\tilde{k}\sqrt{\tilde{k}^2+\etadd\left(2+\edd^{-1}\right)+\etadd\, G(\tilde{k})}\,,
    \label{eqn:spectrum}
\end{equation}
where $\hbar$ is the reduced Planck constant, $\omega_z$ is the axial trap frequency, $\tilde{k}=k\ell_z/\sqrt{2}$, $\ell_z$ is the axial oscillator length, and the function $G(\tilde{k})<0$ captures the reduction in interaction energy arising from head-to-tail dipole alignment.
This spectrum, shown in Fig.~\ref{Fig1}a, exhibits a softening at a wavevector $\krot\sim 1/\ell_z$, and its rotonic character is controlled by two dimensionless quantities. 
First, the dipolar parameter $\edd= a_\textrm{dd}/a_s$~\cite{Koch:2008} determines the strength of long-ranged magnetic forces between atoms relative to the short-ranged van der Waals forces. Here $a_s$ is the $s$-wave scattering length, which is tunable via a Feshbach resonance~\cite{Patscheider:2022}, $a_\textrm{dd}=65.4~a_0$ is the dipolar length of erbium, and $a_0$ is the Bohr radius. 
Second, $\etadd=4\sqrt{2 \pi}\,\add \ell_z n$ compares the dipolar interaction energy scale $\sim\hbar^2 \add n/(m\ell_z)$ to the kinetic energy cost of a rotonic excitation $\sim\hbar^2/(m\ell_z^2)$, where $n$ is the two-dimensional number density and $m$ is the particle mass.
In our experiment, we tune $\etadd$ via $\ell_z$.
We thus anticipate greater rotonic edge density enhancement for larger values of either $\edd$ or $\etadd$.

While $\varepsilon(k)$ determines the density response function $\chi(k)$~\cite{Pitaevskii:2016, Bisset:2019}, a boundary is inherently non-perturbative, meaning that linear response theory is not quantitatively applicable.
In liquid helium, this necessitates density functional approaches~\cite{Dalfovo:1992,Dalfovo:1994}, while here we solve the Gross-Pitaevskii (GP) equation for the ground state density profile~\cite{methods}.  
We show this in Fig.~\ref{Fig1}b for the case of a hard wall and we identify three physically distinct regimes.
The initial growth $n(x)= n_0(m\mu/\hbar^2)x^2$ is fixed by the requirement that the momentum flux into the fluid at its boundary $\hbar^2(\partial \sqrt{n}/\partial x)^2/(2 m )$ equals the bulk thermodynamic pressure $n_0 \mu/2$~\cite{methods}. Here $n_0$ denotes the number density in the bulk. The long-ranged dipolar interactions in the fluid do not affect the initial quadratic
functional form, only its curvature via their contribution
to the chemical potential $\mu$. 
Far from the wall at $x\gg\ell_z$, atoms are insensitive to the precise edge density structure and predominantly experience an effective attraction toward the boundary, owing to the absence of dipolar repulsion from the trap exterior. This results in algebraic decay $n(x)/n_0=1+3\edd\ell_z/[(1+2\edd)\sqrt{2\pi}x]$ of the density toward its bulk value~\cite{methods}.
In the intermediate regime, the fluid exhibits a pronounced enhancement of the density, reflecting a strong rotonic density response to the confining boundary. This is analogous to the long-predicted edge modulation in liquid helium and fractional quantum Hall fluids.

We confine a BEC of $ 50\times10^3$ atoms of $^{166}$Er, spin-polarized in the ground state, in the optical trap shown in Fig.~\ref{Fig1}c. An attractive sheet beam provides an axial harmonic trap with tunable frequency $\omega_z\in 2\pi\times [193,501]~$Hz. 
Sharp confining walls are formed by repulsive laser light at $370~$nm. The spatial intensity profile is controlled using a digital micromirror device imaged onto the atoms via a microscope objective with a numerical aperture of $0.6$~\cite{methods}.
We use this setup to project both circular and annular homogeneous optical traps, each with an outer radius of $8.4~\mu$m.

The cloud is initially prepared with $\edd= 0.9$ and $\etadd=4.2$. We then sweep either $\edd$ or $\etadd$ to its value of interest over $50~$ms and hold for a further $20~$ms before obtaining an absorption image of the axially-integrated atomic density distribution via the same microscope objective used for projecting the optical boundary~\cite{methods}.
In Fig.~\ref{Fig1}d, we show the evolution of the density profile with increasing dipolar character $\edd$. 
Even for the least dipolar samples, a density enhancement is observed at the boundary, as anticipated for a rotonic fluid. This enhancement increases with $\edd$, accompanied by significant density fluctuations throughout the bulk and a growing azimuthal density modulation of the high-density boundary region.

\begin{figure}[t]
    \includegraphics[width=1\linewidth]{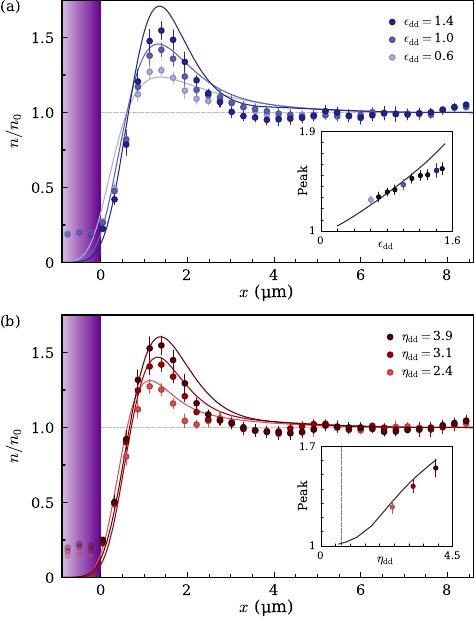}
    \caption{
    \textbf{Density profile at the boundary of a dipolar superfluid.} 
    Data show the observed atomic density distribution $n(x)$ in the vicinity of a confining wall, shown in purple, for different values of \textbf{a)}~the dipolar character $\edd$ at fixed $\etadd=4.2$, and \textbf{b)}~the relative dipolar energy $\etadd$ at fixed $\edd=1.3$. Each profile is obtained from the average of $100$ independent realizations, and is normalized to the bulk density $n_0$. 
    While the initial density growth close to the wall is similar for all curves, the peak density enhancement grows monotonically with both $\edd$ and $\etadd$, consistent with the predicted increase in rotonic character of the fluid. The insets show the variation in the peak density as a function of $\edd$ and $\etadd$, respectively.
    The solid curves in both the main plots and the insets show theoretical predictions obtained by solving the GP equation for the ground state with no free parameters.
    The vertical dashed line in the lower inset indicates the value of $\etadd$ below which $\ell_z$ is smaller than the wall potential lengthscale $\sim 100~$nm and the rotonic response is suppressed~\cite{Juhasz:2022}.
    Error bars indicate the standard deviation of the density across the $2.4~\upmu$m-wide cut from which the profiles are obtained.
    }
    \label{Fig2}
\end{figure}

\begin{figure}[t]
    \includegraphics[width=1\linewidth]{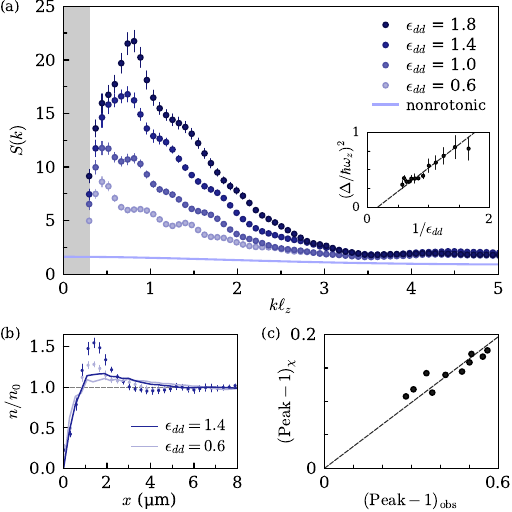}
    \caption{
    \textbf{Density fluctuations in the bulk fluid.} 
    \textbf{a)}~We extract the structure factor $S(k)$ from the power spectrum of density fluctuations. 
    The structure factor exhibits a peak at $k\ell_z\sim 1$ which grows in magnitude with the dipolar character $\edd$, consistent with roton-dominated fluctuations.
    All data significantly exceed the theoretical result for a non-dipolar condensate of the same temperature and chemical potential, shown by the solid line.
    The inset shows the roton energy $\Delta$ obtained from the maximum of $S(k)$, while the dashed line indicates the linear dependence expected for a two-dimensional fluid.
    Error bars are obtained using bootstrap sampling. The shaded gray region indicates the low-$k$ cutoff set by the finite extent of the bulk.
    \textbf{b)}~We use the fluctuation-dissipation theorem at high temperature, $S(k)\propto\chi(k)$, to construct a linear response estimate for the edge density profile, shown by solid lines, while the points indicate the measured profiles.
    \textbf{c)}~We plot the peak edge density enhancement obtained from this linear response model against its measured value, finding an approximately proportional relation, shown by the dashed line. This illustrates the close link between density fluctuations of the bulk and the density response of the edge.
    }
    \label{Fig3}
\end{figure}

To quantify the edge density enhancement, in Fig.~\ref{Fig2} we plot $n$ as a function of distance from the wall $x$ for several values of both $\edd$ and $\etadd$. Since an enhanced static density response is necessarily accompanied by increased fluctuations, we obtain the average density profile over $100$ independent realizations of the fluid. To minimize distortions arising from residual inhomogeneities in the sheet beam, which primarily occur transverse to its propagation direction, we obtain $n(x)$ from a $2.4~\upmu$m-wide cut through the trap center and oriented along the sheet propagation axis.
We observe that the initial growth in density at the boundary is similar for all values of $\edd$ and $\etadd$. This is consistent with our expectation that $n(x)\propto \mu x^2$ at the wall, and we note modest variation in the chemical potential across the parameter ranges addressed~\cite{methods}.
However, the enhancement of $n(x)$ above $n_0$ increases dramatically with both dipolar parameters $\edd$ and $\etadd$, as shown in the insets.
In addition to the experimental data, we show theoretical curves obtained from solving the GP equation to find the ground state density profiles~\cite{methods}, which show good agreement without any free parameters.
The residual inhomogeneity of our trap makes it difficult to distinguish the predicted algebraic decay of the density toward its bulk value, which is at the level of a few percent in its range of validity $x\gg\ell_z$. 

The static density enhancement at the boundary is accompanied by rotonic density fluctuations across the fluid bulk. 
To explore this connection, we characterize these fluctuations by obtaining the static structure factor $S(k)=\langle | \delta n_\textbf{k} |^2\rangle/N_\textrm{b}$ from the bulk for several values of $\edd$~\cite{methods}. Here $\delta n_\textbf{k}$ is the Fourier transform of the deviation $\delta n(\textbf{r})$ of the density from its mean value, the atom number inside the analysed region is $N_\textrm{b}$, and $\langle \dots \rangle$ denotes an average over many realizations of the fluid. 
The structure factor is plotted in Fig.~\ref{Fig3}a and exhibits a peak at a wavevector $\sim 1/\ell_z$ which grows monotonically with $\edd$.
For comparison, we calculate $S(k)$ for a non-dipolar condensate at the same temperature as and with the mean chemical potential of the dipolar fluids~\cite{methods}. The resulting curve is shown by the solid line and lies significantly below the data, illustrating the dramatic effect that rotonic softening has upon density fluctuations.

Depending on the ratio of thermal energy $\kb T$ to $\varepsilon(k)$, where $\kb$ is the Boltzmann constant and $T$ is the temperature, density fluctuations arise from either thermally populated collective excitations or their associated zero-point energy. For a weakly-interacting BEC~\cite{Pitaevskii:2016,Hertkorn:2021},
\begin{equation}
    S(k)=\frac{\hbar^2 k^2}{2 m \varepsilon(k)}\coth\left(\frac{\varepsilon(k)}{2\kb T}\right).
    \label{eq:sk}
\end{equation}
Direct thermometry of the fluid is challenging owing to the absence of any detectable thermal atoms, and we therefore extract the temperature $T = 44(5)~$nK from the high-$k$ behaviour of $S(k)$~\cite{methods}. 
Since the roton wavevector is $1/\ell_z$, the low-$k$ phononic dispersion $\varepsilon\approx \hbar k\sqrt{\mu/m}$ gives a roton energy $\Delta\lesssim\sqrt{\mu \hbar \omega_z} \sim \kb \times 10~$nK. This implies that the thermal energy greatly exceeds the roton energy in our system and that the fluctuations characterized in Fig.~\ref{Fig3}a are primarily thermal. We therefore use the high-temperature relation $S(k)\approx\kb T \hbar^2k^2/(m\varepsilon^2(k))$ to extract $\Delta$ from the peak value of $S(k)$. This is plotted in the inset of Fig.~\ref{Fig3}a, and we find an approximately linear dependence of $\Delta^2$ upon $1/\edd$ as expected from the form of Eq.~(\ref{eqn:spectrum}). 

The density fluctuations of a fluid are intimately linked to its density response via the fluctuation-dissipation theorem $S(k)\geq \kb T \chi(k)$, which becomes an equality in the limit $\kb T\gg\varepsilon(k)$~\cite{Pitaevskii:2016,Bisset:2019}.
This is satisfied by our experiment for $k\ell_z\lesssim \sqrt{2\kb T/(\hbar\omega_z)}\approx 3$, meaning that in this range the structure factor plotted in Fig.~\ref{Fig3}a is approximately proportional to $\chi(k)$. 
We therefore use our measurement of $S(k)$ from the bulk to construct a linear response estimate of the density structure at the boundary, by following~\cite{Regge:1972} and assuming that the modulation is proportional to that arising from a delta function perturbation. This yields $n(x)/n_0=1-\int\textrm{d}k\chi(k)\exp(i k x)/\int\textrm{d}k\chi(k)$.
The resulting profiles, examples of which are shown in Fig.~\ref{Fig3}b, qualitatively capture the initial rise, peak position, and decay toward the bulk, but systematically underestimate the observed peak value.
In Fig.~\ref{Fig3}c we plot the observed and linear response peak density enhancements against each other, finding an approximately proportional relationship and illustrating the link between the static density response of the fluid at the edge and its transient density fluctuations in the bulk.

\begin{figure}[t]
    \includegraphics[width=1\linewidth]{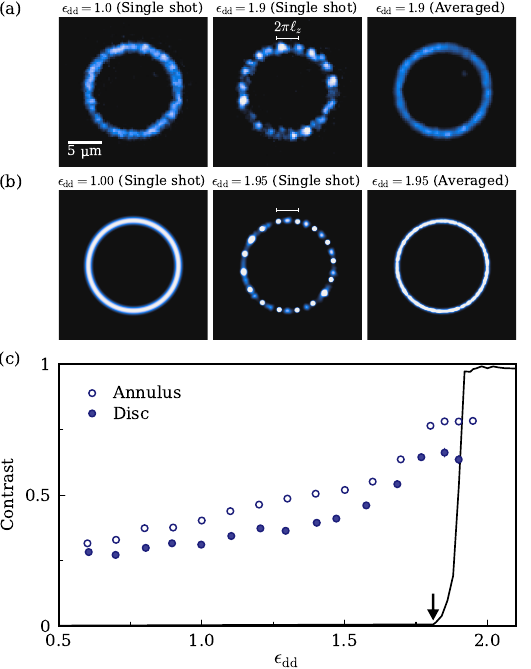}
    \caption{\textbf{Rotonic density modulation of an annular dipolar fluid.} 
    \textbf{a)}~We confine a dipolar condensate in an annular trap, whose evolution is similar to the boundary of the two-dimensional fluid shown in Fig.~\ref{Fig1}d.
    As $\edd$ rises, we observe density modulation at the rotonic lengthscale $2\pi\ell_z$, indicated by a white scale bar. 
    The modulation phase is random, as evidenced by the azimuthal symmetry of an average image.
    \textbf{b)}~We simulate our experiment via real-time evolution of the GP equation, and obtain similar density profiles to those observed in the experiment.
    \textbf{c)}~To compare the evolutions of the annular fluid and the boundary of the disc-shaped cloud, we plot the contrast of their one-dimensional azimuthal density.
    This shows similar behaviour for both trapping geometries, reflecting the quasi-one-dimensional nature of the boundary of the two-dimensional fluid.
    The contrast of the zero-temperature simulated cloud is shown by a solid line, which exhibits a sudden onset. 
    This is associated with a change in the ground state from an azimuthally-symmetric condensate to a droplet phase, indicated by an arrow.
    The non-zero temperature in the experiment results in growing thermal fluctuations on approaching this instability.
    Error bars indicate standard error of the mean.
    }
    \label{Fig4}
\end{figure}

In addition to the increasing magnitude of edge density enhancement with dipolar character, for $\edd\gtrsim1.6$ we also observe significant azimuthal density modulation of the fluid's boundary. 
This has a lengthscale $\sim 2\pi\ell_z$, suggesting softening of a rotonic mode with a wavevector oriented along the quasi-one-dimensional edge, as discussed in recent theoretical work~\cite{Roccuzzo:2022}.
Recently, the possibility of an analogous edge roton was predicted in the context of fractional quantum Hall fluids~\cite{Jolad:2010}.

To explore this similarity, we compare the growth in density modulation of the boundary region with that of an annular fluid, illustrative images of which are shown in Fig.~\ref{Fig4}a. The outer radius of $8.4~\upmu$m, inner radius of $6.3~\upmu$m, and atomic number density are chosen to closely match the density-enhanced boundary region of Fig.~\ref{Fig1}d and we perform the same experimental sequence as for the disc-shaped cloud before.  
As $\edd$ is increased, we observe a pronounced growth in density modulation with a lengthscale that closely matches the rotonic expectation $2\pi\ell_z$. 
This modulation is not present in the average image obtained from $30$ iterations of the experiment, indicating that its phase is random.
We also simulate our experiment via real-time evolution under the GP equation and show illustrative images in Fig.~\ref{Fig4}b, which closely match the observed profiles.
While modulation of the simulated fluid appears at slightly higher values of $\edd$ than in the experiment, we find that this is sensitive to the atom number and wall intensity profile of the confining annulus.

To quantify the growth in density modulation with $\edd$, we obtain the azimuthal one-dimensional density close to the edge via radial integration, and in Fig.~\ref{Fig4}c we plot the density contrast for the annular fluid, the boundary of the two-dimensional cloud, and the GP real-time simulations~\cite{methods}. 
The two trapping geometries exhibit similar behaviour, consistent with the azimuthal density modulation of the boundary arising from a quasi-one-dimensional rotonic mode.
The growth in contrast of our zero-temperature simulated fluid is more sudden, and its onset coincides with a change in the underlying ground state from an azimuthally-symmetric condensate to isolated droplets, indicated by an arrow.  
This indicates that the modulation observed under real-time evolution of the GP equation corresponds to a dynamical instability associated with an azimuthal rotonic mode, in analogy to experiments in harmonically-trapped gases~\cite{Chomaz:2023,Recati:2023}. 
In the experiment, the non-zero temperature necessarily results in substantial thermal density fluctuations as we approach this instability, which makes the growth in contrast more gradual.

These experiments directly realize a long-standing prediction of density modulation at the boundary of a rotonic quantum fluid, and provide a canonical example of emergent structure arising purely from interparticle interactions and a confining potential.
Our annular fluid natively hosts periodic boundary conditions and is thus well suited for studying hydrodynamics and excitations~\cite{Sindik:2024,Hertkorn:2024,Platt:2025}. 
More broadly, the nascent availability of uniform potentials for dipolar quantum gases will be beneficial for exploring collective excitations, transport, correlations, turbulence, and critical phenomena.

\begin{center}
    \textbf{Acknowledgments}
\end{center}
We thank Jens Hertkorn and Martin Zwierlein for useful discussions and critical comments on the manuscript, and Jared Rivera for early contributions to the repulsive potential setup.
This work was supported by the NSF through the Center for Ultracold Atoms (PHY-2317134)
and Grant PHY-2207367,
the AFOSR Young Investigator Program (FA9550-22-1-0066),
and the David and Lucile Packard Foundation (2023-76156).
This material is based upon work supported by the
Defense Advanced Research Projects Agency (DARPA) under Agreement No. HR00112490527.
J.K. acknowledges support from  the National Science Foundation Graduate Research Fellowship under Grant No. 2141064.
K.K. acknowledges support from the MIT Quantum Initiative postdoctoral fellowship program funded by the Gordon and Betty Moore Foundation through Grant No. GBMF13896.

\begin{center}
    \textbf{Author contributions}
\end{center}
J.K., K.K., and W.S.H. performed the measurements. 
J.K. and K.K. analysed the data. 
R.E., K.K., and A.P. performed the numerical simulations.
R.J.F. developed the theoretical description. 
K.K. and R.J.F. supervised the project. 
All authors contributed to the interpretation of the results and the preparation of the manuscript.

\begin{center}
    \textbf{Competing interests}
\end{center}
The authors declare no competing interests.

\bibliography{refs}

\begin{center}
    \textbf{Methods}
\end{center}

\paragraph{Uniform potential} The strong $401~$nm transition in erbium means that engineering repulsive optical potentials necessitates laser wavelengths below this value. To provide sharp confining walls, we use light at $370~$nm generated by a tunable Ti:Sapph laser at $740~$nm followed by a doubling cavity. This light is used to illuminate a digital micromirror device (DMD), which is then imaged onto the atoms using a microscope objective. The objective numerical aperture $\textrm{NA}=0.6$ yields an edge spread function whose theoretical form is well-captured by a sigmoid function $\sim 1/(1+\exp(-x/d))$ with width $d=90~$nm. The pixel pitch of the DMD is $13.6~\upmu$m, which corresponds to $160~$nm at the atoms, below our imaging resolution.

To estimate the inhomogeneity of the confining potential, we extract the density variation of a minimally dipolar sample with $\edd=0.6$, shown in Fig.~\ref{Fig1}c. First, we obtain an average density profile from $100$ sequential realizations. Next, we construct a histogram of the density measured at each pixel normalised to the mean density. We exclude the outer $3~\upmu$m of the cloud from this analysis as it exhibits rotonic density enhancement even for the lowest accessible $\edd$. The histogram shows a standard deviation of $0.07$, bounding the density inhomogeneity arising from trap imperfections. From solution of the GP equation for our experimental parameters, we obtain a chemical potential of $\approx h\times 770~$Hz for the fluid shown in Fig.~\ref{Fig1}c, meaning that our trapping potential exhibits a variation of $\sim h\times 50~$Hz across the sample. 

\paragraph{Chemical potential of the dipolar fluid} The chemical potential is inferred from imaginary time propagation of the GP equation to find the ground state of our dipolar fluid, and varies with both $\edd$ and $\etadd$. For the weakest axial confinement of $\omega_z=2\pi\times 193~$Hz, the chemical potential varies from $h\times 870~$Hz for $\edd=0.6$ to $h\times 650~$Hz for $\edd=1.6$, while for varying axial trapping at fixed $\edd=1.3$ we find $h\times 690~$Hz for $\omega_z=2\pi\times 193~$Hz and $h\times 1500~$Hz for $\omega_z=2\pi\times 501~$Hz. The values given here include the zero-point energy of the axial harmonic trap.

\paragraph{Imaging of the gas} Measurement of the atomic density distribution is performed via high-intensity absorption imaging. The imaging light has a saturation parameter of $\sim 20$ and a pulse duration of $\tau=2~\upmu$s, resulting in each $^{166}$Er atom undergoing axial motion of $ 2 \pi\Gamma \hbar \tau^2 /(4 m\lambda)=1.1~\upmu$m during the pulse, which is comparable to the imaging depth of focus. Here $\Gamma=2\pi\times 30~$MHz is the natural linewidth of the blue imaging transition with a wavelength $\lambda=401~$nm. The transverse diffusion of an atom during the imaging pulse is $4\pi\sqrt{\Gamma/2}\hbar \tau^{3/2}/(3m\lambda)\approx 110~$nm, which is much smaller than the optical resolution.

\paragraph{Atom number calibration} For a parameter-free comparison between our measured rotonic density modulation and the predictions obtained by solving the GP equation, we calibrate our absolute atom number. This is done via the onset of BEC in a thermal gas held in a cylindrically-symmetric dipole trap formed by two crossed Gaussian beams. First, we measure the critical atom number via low-intensity absorption imaging after time-of-flight. Next, we calculate the theoretical critical number using our known temperature, trap frequencies, scattering length, and dipolar length. We incorporate interaction-induced shifts to the critical number~\cite{Smith:2011,Glaum:2007,Krstajic:2025}
\begin{align}
    &N_c=A\zeta(3)\left(\frac{\kb T}{\hbar \bar{\omega}}\right)^3\nonumber\\
    &\times\left(\frac{1}{1-3.426\left\{\frac{a_s}{\lambda_{\textrm{dB}}}-\frac{\add}{\lambda_{\textrm{dB}}}f\left(\frac{\omega_z}{\omega_r}\right)\frac{3\cos^2\theta-1}{4}\right\}+46\left(\frac{a_s}{\lambda_{\textrm{dB}}}\right)^2}\right)^3,
\end{align}
where $\theta=0$ is the angle between the magnetic dipoles and trap symmetry axis, $\lambda_{\textrm{dB}}$ is the thermal de Broglie wavelength, $\bar{\omega}=(\omega_r^2\omega_z)^{1/3}$,  $\omega_{r,z}$ are the radial and axial trap frequencies, the function $f(x)$ is given in~\cite{Glaum:2007}, and the prefactor $A\approx 1.09$ accounts for the non-harmonic profile of the Gaussian trapping beams ~\cite{Campbell:2012}. We find a ratio of imaged to calculated atom number of $0.73$ for our low-intensity imaging, which is used to obtain all atom numbers given in the paper.

\paragraph{Measurement of the structure factor} We characterize the density fluctuations of the bulk fluid by obtaining the static structure factor $S(k)=\langle | \delta n_\textbf{k} |^2\rangle/N_\textrm{b}$ where $\delta n_\textbf{k}$ is the Fourier transform of the deviation $\delta n(\textbf{r})$ of the density from its mean value, and $N_\textrm{b}$ is the atom number inside the analysed region. 
We obtain $100$ images from identical experimental iterations and discard any for which the atom number deviates by $>10~\%$ from the mean value, typically fewer than 10 images.
We construct the average density profile $\bar{n}(\textbf{r})$. Each image thus yields a corresponding fluctuation profile $\delta n(\textbf{r})=n(\textbf{r})-\bar{n}(\textbf{r})$. We then multiply by a circular top-hat function of radius $6~\upmu$m to exclude the density-enhanced boundary region from our analysis, and remove any constant offset or linear gradients from $\delta n(\textbf{r})$, which correspond to atom number fluctuations and centre-of-mass sloshing. We obtain the power spectrum $|\delta n_\textbf{k}|^2=|\int\textrm{d}^2\textbf{r}\exp(i\textbf{k}\cdot\textbf{r})n(\textbf{r})|^2$, divide by the number of atoms $N_\textrm{b}$ in the bulk region, and average over all images to obtain $S(\textbf{k})$. Finally, we perform an azimuthal average of $S(\textbf{k})$ in momentum space to yield $S(k=|\textbf{k}|)$ which is plotted in Fig.~\ref{Fig3}a of the main text. To incorporate the effect of finite imaging resolution, we correct the measured $S(k)$ by the theoretical modulation transfer function according to $S(k)\rightarrow S(k)/\textrm{MTF}^2(k)$ where 
\begin{equation}
\textrm{MTF}(k)= 
\frac{2}{\pi}\left\{\cos^{-1}\left(\frac{k}{k_\textrm{c}}\right) - \left(\frac{k}{k_\textrm{c}}\right)\sqrt{1-\left(\frac{k}{k_\textrm{c}}\right)^2}\right\}\nonumber.
\end{equation}
Here the wavevector $k\leq k_\textrm{c}=4\pi\textrm{NA}/\lambda$, $\textrm{NA}=0.6$ is the numerical aperature of our microscope objective, and $\lambda=401~$nm is the imaging wavelength ~\cite{Smith:2008}.

\paragraph{Thermometry of the gas} Direct measurement of the fluid's temperature is challenging, since there is no detectable thermal population of atoms under time-of-flight imaging. To estimate the temperature, we note that in the high-$k$ limit the dispersion takes the form $\varepsilon(k)\rightarrow {\hbar^2 k^2}/({2m})$. 
 Following Eq.~(\ref{eq:sk}) of the main text, we plot $\textrm{coth}^{-1}S$ as a function of $(k\ell_z)^2$ and anticipate a linear relation with slope $\hbar\omega_z/(4\kb T)$. 
This is shown in Fig.~\ref{FigS1} and we extract the temperature via a linear fit in the region $k\ell_z\gtrsim 2$, which is plotted in the inset as a function of $\edd$. The mean and standard deviation yield a temperature estimate $44(5)~$nK. 

\setcounter{figure}{0}
\renewcommand{\thefigure}{S\arabic{figure}}
\begin{figure}[t]
    \includegraphics[width=1\linewidth]{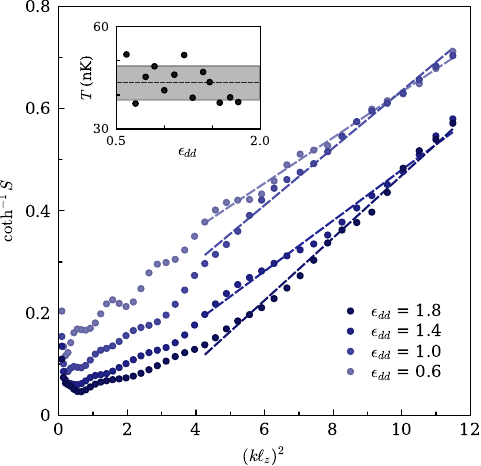}
    \caption{
    \textbf{Estimating temperature from the structure factor.} 
    In the high-$k$ limit, the dispersion relation of a dipolar condensate becomes free particle-like, resulting in a linear relation between $\coth^{-1}S$ and $k^2$. We fit the data for $k\ell_z \gtrsim 2$, shown by dashed lines, and from the slope extract the temperature. This is shown in the inset as function of $\edd$ with a mean and standard deviation $44(5)~$nK.
    }
    \label{FigS1}
\end{figure}

\paragraph{Structure factor of a non-dipolar condensate}
In Fig.~\ref{Fig3}a we compare our measured $S(k)$ with the theoretical curve for a non-dipolar condensate, with the same temperature $T=44~$nK as the dipolar fluids addressed in the experiment and with the same chemical potential $\mu=h\times 640~$Hz as the $\edd=1$ sample. Here we have subtracted the zero-point energy associated with the axial motion from the chemical potential. The structure factor is obtained using Eq.~(\ref{eq:sk}) of the main text, into which we substitute the dispersion of a non-dipolar BEC $\varepsilon(k)=\sqrt{\hbar^2 k^2/(2m)}\sqrt{\hbar^2 k^2/(2m)+2\mu}$.

\paragraph{Theoretical edge structure of a dipolar superfluid}
In Fig.~\ref{Fig1}b of the main text, we show the two-dimensional number density $n(x)$ of a dipolar superfluid in the vicinity of a hard-wall potential located at $x=0$. The profile is obtained via numerical solution of the GP equation, and we also show analytic forms capturing the small- and large-distance limiting behaviours which are derived below.

The initial growth in the density is obtained by noting that the momentum flux into the fluid from the confining potential must equal the bulk thermodynamic pressure $n_0\mu/2$. The momentum flux is obtained by integrating the three-dimensional GP equation along the confined $z-$direction and assuming the system is translationally invariant along $y$, yielding
\begin{align}
\psi''(x)&=\frac{2m}{\hbar^2}\left(U(x)+\phi(x)-\mu\right)\psi(x)\nonumber\\
&=\lambda(x)\psi(x)
\label{Eq:GP}
\end{align}
where prime denotes differentiation with respect to $x$, $U(x)$ is the confining potential, and $\phi(x)$ is the potential experienced by an atom at position $x$ due to interactions with other atoms. Since we are considering the ground state, the wavefunction $\psi(x)$ is real and $n(x)=\psi^2(x)$. 

Noting that the total force density on the fluid is $-n(U'+\phi')$, we identify the momentum flux $\sigma$ from the relation $\sigma'=-n(U'+\phi')$. Employing Eq.~(\ref{Eq:GP}) we find
\begin{align}
    -n(U'+\phi')&=\frac{-\hbar^2}{2m}\lambda'\psi^2=\frac{-\hbar^2}{2m}(\psi \psi'''-\lambda\psi\psi')\nonumber\\
    &=\frac{-\hbar^2}{2m}(\psi \psi'''-\psi'\psi'')\nonumber\\
    &=\frac{\partial}{\partial x}\left[\frac{\hbar^2}{2m}(\psi'^2-\psi\psi'')\right]
\end{align}
from which we identify the momentum flux $\sigma=\hbar^2(\psi'^2-\psi\psi'')/2m$. 

Since the wavefunction must vanish at the boundary, it varies as $\psi(x)\approx \gamma x$, giving a momentum flux $\hbar^2\gamma^2/2m$ into the fluid at the wall location. Equating this to the bulk thermodynamic pressure $n_0\mu/2$, we find $\gamma^2=m\mu n_0/\hbar^2$ yielding a density profile
\begin{equation}
    \lim_{x\to 0}\frac{n(x)}{n_0}=\frac{m\mu}{\hbar^2}x^2.
\end{equation}
We note that this initial quadratic evolution is identical to that of a condensate with only contact interactions, where the edge structure takes the analytic form $n(x)=n_0\tanh^2(\sqrt{\frac{m\mu}{\hbar^2}}x)$. The inclusion of long-ranged dipolar interactions does not affect the initial quadratic functional form, only its curvature via their contribution to the chemical potential. For the quadratic profile plotted in Fig.~\ref{Fig1}b of the main text, we use the chemical potential obtained from solving the GP equation and subtract the contribution from the zero-point axial motion, which is common to all atoms.

Far from the confining boundary, we anticipate that the atoms respond primarily to an effective long-ranged attraction toward the edge, owing to the absence of dipolar repulsion from the empty region. Using the interaction potential energy of two side-by-side dipoles $3 g_s \edd / (4\pi r^3)$, where $r$ is the interatomic distance, we integrate over the empty region $x<0$ to obtain an effective attractive potential
\begin{align}
    Q(x)&=n_0 \frac{3g_s\edd}{4\pi}\int_{-\infty}^\infty\textrm{d}y\int_{-\infty}^x\textrm{d}s\frac{1}{(s^2+y^2))^{3/2}},\nonumber\\
    &=\frac{3g_s\edd}{2\pi}n_0\frac{1}{x}.
\end{align}
From the compressibility of a two-dimensional dipolar fluid $\partial n/\partial \mu=l_z \sqrt{2\pi} / (g_s(1+2\edd))$~\cite{Baillie:2015} we thus infer a far-field density profile
\begin{align}
    \lim_{x\to\infty}\frac{n(x)}{n_0}&=1+\frac{\partial n}{\partial \mu}\frac{Q(x)}{n_0}\notag\\
    &=1+\frac{3\edd}{1+2\edd}\frac{1}{\sqrt{2\pi}}\frac{l_z}{x},
\end{align}
which is shown in Fig.~\ref{Fig1}b with no free parameters.

\paragraph{Excitation spectrum} The spectrum of collective excitations for a planar, uniform dipolar condensate has been addressed in several theoretical works~\cite{Santos:2003,Baillie:2015,Fischer:2006}. To gain intuition into how the rotonic behaviour emerges and scales, we consider the case that the fluid occupies only the axial ground state, yielding a spectrum
\begin{align}
    \varepsilon(k)=&\sqrt{\frac{\hbar^2 k^2}{2m}}\notag\\\times&\sqrt{\frac{\hbar^2 k^2}{2m}+\sqrt{\frac{2}{\pi}}\frac{n g_s}{\ell_z}\left[1+\edd\left(2+ G\left(\frac{k\ell_z}{\sqrt{2}}\right)\right)\right]}
    \label{eqn:Dispersion}
\end{align}
where $G(x)=-3\sqrt{\pi}x\exp(x^2)\textrm{erfc}(x)$ and $\textrm{erfc}(x)$ denotes the complementary error function. Introducing dimensionless variables $\tilde{k}=k\ell_z/\sqrt{2}$, $\edd= a_\textrm{dd}/a_s$ and $\etadd=4\sqrt{2 \pi}\,\add \ell_z n$, the spectrum takes the form given in Eq.~(\ref{eqn:spectrum}) of the main text,
\begin{equation}
    \frac{\varepsilon(k)}{\hbar\omega_z}=\tilde{k}\sqrt{\tilde{k}^2+\etadd\left(2+\edd^{-1}\right)+\etadd\, G(\tilde{k})}\,,
\end{equation}

Physically, the first term within the square root corresponds to the kinetic energy cost of an excitation at momentum $\tilde{k}$, the second term gives a constant energy cost arising from the zero-momentum components of both contact and dipolar interactions, and the third term describes attraction between neighbouring dipoles which softens the energy cost of exciting a density modulation.

In our experiment, the chemical potential is $\sim 4\hbar\omega_z$ meaning the condensate wavefunction contains an admixture of higher axial states. However, we anticipate our predictions regarding the role of $\edd$ and $\etadd$ in quantifying rotonic character to remain valid. 
\paragraph{Quantifying density modulation via density contrast}
In Fig.~\ref{Fig4}c we compare the azimuthal density modulation of the annular fluid with the boundary of the two-dimensional cloud. 
We integrate the density over the radial interval $[6.8~\upmu\textrm{m},7.9~\upmu\textrm{m}]$ to construct the azimuthal density profile $n_\textrm{1D}$ and remove Fourier components corresponding to azimuthal periods of $2\pi$ and $\pi$ to mitigate effects of residual trap inhomogeneity and cloud sloshing.
The contrast is defined as $(\max\{n_{1D}\}-\min\{n_\textrm{1D}\})/(\max\{n_\textrm{1D}\}+\min\{n_\textrm{1D}\})$.
\paragraph{Gross-Pitaevskii equation}
The dipolar fluid at zero temperature is described by the extended Gross-Pitaevskii equation
\begin{align}
    i\hbar \frac{\partial}{\partial t} \psi (\mathbf{r},t) = &\left[-\frac{\hbar^2\nabla^2}{2m}+U(\mathbf{r})+g_s|\psi(\mathbf{r},t)|^2 + \gamma |\psi(\mathbf{r},t)|^3\right.\nonumber\\
     &\left.+\int d\mathbf{r'}V_\textrm{dd}(\mathbf{r}-\mathbf{r'})|\psi(\mathbf{r'},t)|^2\right]\psi(\mathbf{r},t),
\label{eqn:GP}
\end{align}
where the dipolar potential is $V_\textrm{dd}(\mathbf{u}) = 3g_s \edd(1-3\cos^2\theta) / (4\pi |\textbf{u}|^3)$, $\theta$ is the angle between $\textbf{u}$ and the dipole polarization axis $\hat{\textbf{z}}$, the contact interaction strength is $g_s = {4\pi\hbar^2 a_s}/{m}$, and the Lee-Huang-Yang correction to the energy is included at the mean-field level via $\gamma = {32}g_s\sqrt{{a_s^3}/{\pi}}\left(1+{3}\edd^2/2\right)/3$ 
~\cite{Lee:1957, Schutzhold:2006, Lima:2011}. The trap potential in cylindrical coordinates is modeled as
\begin{equation*}
    U(\mathbf{r})=\frac{U_0}{1+e^{(\rho_{\mathrm{out}}-\rho)/w(z)}}+\frac{1}{2}m\omega_z^2z^2,
\end{equation*}
corresponding to a repulsive optical wall potential of radius $\rho_\textrm{out}$, height $U_0$, and characteristic width $w$. 
To obtain $w$, we calculate the diffraction-limited edge spread function $\mathrm{ESF}(x) = \int_{-\infty}^x dx'\int_{-\infty}^{\infty} dy'\mathrm{PSF}(x',y')$ where $\textrm{PSF}(x,y)$ denotes the point spread function of our microscope objective. We then fit a sigmoid function to the ESF to obtain the in-focus width $w$.
To reproduce the experimental conditions, we calibrate the trap depth $U_0$ by comparing the optical power at which the condensate becomes unconfined to the chemical potential obtained via imaginary-time propagation of the GP equation. 

\begin{figure}[t]
    \includegraphics[width=1\linewidth]{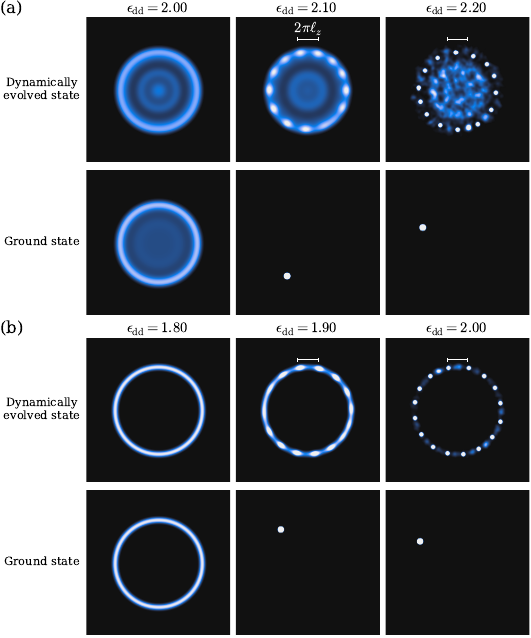}
    \caption{
    \textbf{Variation of the ground and dynamically evolved states with dipolar character.} 
    We show the density distribution obtained for both real-time evolution of the GP equation following our experimental protocol and imaginary-time evolution. We perform simulations for both the \textbf{a)}~disc and \textbf{b)}~annular traps used in the experiment.
    We observe a sharp crossover in the ground state geometry, from an azimuthally-symmetric condensate to an isolated droplet. 
    Under real-time evolution, this change in the ground state is heralded by the emergence of periodic density modulation at the rotonic lengthscale $\sim2\pi\ell_z$.
    The simulated atom number is $48\times 10^3$ in the disc trap and $23\times 10^3$ in the annular trap, chosen to match the experimentally measured atom numbers. 
    }
    \label{FigS2}
\end{figure}

We also account for the experimentally calibrated defocus of $2.8~\upmu$m of the $370~$nm focal plane from the $401~$nm focal plane of the objective.

To determine the energetic ground state for a given set of experimental parameters, we perform imaginary-time evolution of Eq.~(\ref{eqn:GP}) via the split-step Fourier method. 
We also perform simulations of the condensate's evolution under our specific experimental sequence. We first determine the ground state at $\edd = 0.87$, at which the condensate in our experiment is initially prepared and loaded into the uniform optical trap. We then perform real-time evolution of the obtained initial state under the same ramp in $\edd$ as in the experiment.

In Fig.~\ref{FigS2}, we compare this dynamically evolved state with the energetic ground state at several values of $\edd$. For low values of $\edd$, the ground and dynamically evolved states closely resemble each other for both the disc and annular trap. However, beyond a critical value of $\edd$, a clear distinction emerges. The ground state exhibits an azimuthal symmetry breaking into a droplet state, while the dynamically-evolved fluid exhibits a pronounced azimuthal density modulation at the rotonic lengthscale $\sim 2\pi\ell_z$. For our measured atom number we find a ground state consisting of a single droplet, while larger atom numbers result in multiple droplets forming a one-dimensional array along the boundary as in~\cite{Roccuzzo:2022}. Furthermore, while the critical value of $\edd$ for the simulated fluid is somewhat larger than that observed in the experiment, we find that this critical point is very sensitive to the simulated atom number. 
This concurrence of the change in ground state geometry with the emergence of density modulation under real-time evolution suggests that the latter should be interpreted as a dynamical instability, associated with the energy of a roton propagating along the azimuthal direction becoming imaginary. Such an instability has been extensively studied in harmonically-trapped systems~\cite{Chomaz:2023,Recati:2023}. 
\newpage
\begin{center}
    \textbf{Data availability}
\end{center}
All data files are available from the corresponding author upon request. 

\begin{center}
    \textbf{Code availability}
\end{center}
The analysis code is available from the corresponding author upon reasonable request.

\end{document}